\documentstyle[12pt]{article}
\begin{document}
\begin{center}
\Large
{\bf
EFFECTIVE ACTION OF COMPOSITE FIELDS FOR GENERAL
GAUGE THEORIES IN  BLT--COVARIANT FORMALISM}

\vspace{0.5cm}
P.M.LAVROV\footnote{e-mail address: lavrov@tspi.tomsk.su},
S.D.ODINTSOV\footnote{e-mail address: sergei@ecm.ub.es}\\
{\it Department of Mathematical Analysis,
Tomsk State Pedagogical University, Tomsk, 634041, Russia
and
    Dept. ECM, Fac.de Fisica, Universidad de Barcelona,
Diagonal 647, 08028 Barcelona, Spain}\\
and\\
A.A.RESHETNYAK\footnote{e-mail address:reshet@phys.tsu.tomsk.su}\\
{\it Quantum Field Theory Department, Tomsk State University, Tomsk,
634050, Russia}\\

\end{center}
\vspace{.5cm}
\large
\begin{quotation}
\normalsize
\noindent{\bf Abstract}

The gauge dependence of the effective action of composite fields for
general gauge theories in the framework of the quantization method
by Batalin,
Lavrov and Tyutin is studied. The corresponding Ward identities are
obtained. The variation of composite fields effective action is
found in terms of new set of operators depending on composite field.
The theorem of the on-shell gauge fixing independence for the
effective action of composite fields in such formalism is proven.
Brief discussion of gravitational-vector induced interaction for
Maxwell theory with composite fields is given.

\end{quotation}

\newpage
\section{INTRODUCTION}
\hspace*{\parindent}

The advanced methods of covariant quantization for general gauge
theories are based either on the BRST symmetry principle realized
in the well-known quantization scheme by Batalin and Vilkovisky [1]
or on the extended BRST symmetry principle recently realized within
the quantization method by Batalin, Lavrov and Tyutin (BLT) [2].
The
various aspects and properties of the gauge field theory within
the BV quantization have been under study for quite a long time
by now and may be considered as well-known ones (see, for example,
reviews [3,4]). On the same time
                the study of properties as well as various possibilities
of interpretation  and generalizations of gauge theories in the BLT
quantization [2] has been started quite recently [5-16].
Following the line of the research of refs.[5-16] present paper
                           is devoted to the study of one of the
central problems arising in     quantum gauge field theory within the
Lagrangian formalism, i.e. gauge dependence of generating functionals
of Green's functions in general gauge theories with composite fields.

Our interest in consideration of composite fields within the BLT-
formalism is caused by
                 to a number of reasons. First of all, since the work
[17] (for a review, see [18]) where the formalism to study the effective
action (EA)
       for composite fields has  been introduced
such EA is often used to
discuss the dynamical chiral symmetry breaking phenomenon in different
models using for example Schwinger-Dyson equations.
                                          Second, in four - fermion
models [19] the fermions form the composite boundstates which may
play the role of Higgs field for discussion of dynamical symmetry
breaking in the Standard Model (see [20] and references therein). Third,
in the models of inflationary Universe the composite boundstate may
play the role of the inflaton. Finally, the Wilson effective action for
composite fields maybe extremely important in recent studies on the
exact results in SUSY theories ( for a review, see [21]).

The paper is organized as follows. In Sec.2 we give a short review of
the BLT formalism. In Sec.3 we derive the Ward identities for
effective action of composite fields in general gauge theories in the
framework of the quantization method by Batalin, Lavrov and Tyutin.
Sec.4 is devoted to the study of
                gauge dependence structure for composite fields EA.
                            In Sec.5 the example of this effective
action is considered. Concluding remarks are given in Sec.6. In our
paper we use the notations of refs.[2].

\section{BATALIN - LAVROV - TYUTIN QUANTIZATION}
\hspace*{\parindent}

In this section we give          a short review of the main features of
BLT--quantization method for general gauge theories.
In order to do this we start from the
                          definition of general gauge theories.

         Let us consider the theory of fields
$A^i(i=1,2,...,n,\varepsilon(A^i) =\varepsilon _i$)for which the
initial classical action ${\cal S}(A)$ is invariant under the gauge
transformations $\delta A^i=R^i_{\alpha}(A) \xi^{\alpha}$:
$$
{\cal S}_{,i}(A) R^i_{\alpha}(A)=0,
$$
$$ \alpha=1,2,...,m, 0<m<n,\;\;
\varepsilon(\xi^{\alpha}) =\varepsilon({\alpha}),\eqno(1)
$$
where     $\varepsilon(\xi^{\alpha})$ are arbitrary functions, and the
$R^i_{\alpha}(A)$ are generators of gauge transformations. We suppose
     the set $R^i_{\alpha}(A)$ being the linearly independent
(case of
irreducible theories) and complete.  One can say that as a consequence
of the condition of completeness the algebra of generators has the
following general form:
$$
R^i_{\alpha ,
j}(A)R^j_{\beta}(A)-(-1)^{\varepsilon_{\alpha}\varepsilon_
{\beta}}R^i_{\beta ,j}(A)R^j_{\alpha}(A)=
$$
$$
-R^i_{\gamma}(A)F^{\gamma}_
{\alpha\beta}(A)-{\cal S}_{,i}(A)M^{ij}_{\alpha\beta}(A),
\eqno(2)
$$
where $M^{ij}_{\alpha\beta}$ satisfy   the conditions
$$
M^{ij}_{\alpha\beta} = -(-1)^{\varepsilon_i\varepsilon_j}
M^{ji}_{\alpha\beta} =
-(-1)^{\varepsilon_{\alpha}\varepsilon_{\beta}}M^{ij}_{\beta\alpha}
$$
The gauge theories whose generators satisfy Eq.(2) are called general
gauge theories. As it
                   has already been mentioned, covariant quantization
of such theories (as well as reducible ones) in the framework of a
standard BRST symmetry in modern form has been proposed by Batalin and
Vilkovisky [1].

To construct the BLT--quantization scheme it is necessary to
introduce the
total configuration space $\phi^A$. For irreducible theories the total
configuration space $\phi^A$ has the following form
$$
\phi^A = (A^i,\;B^{\alpha},\;C^{\alpha a}),\;\;\;
\varepsilon (\phi^A) = \varepsilon_A.
$$
Here $C^{\alpha a}$ is      $Sp(2)$-- doublet of ghost ($a =1$) and
antighost ($a = 2$) fields (Faddeev-Popov fields), $B^{\alpha}$ are
auxiliary fields
$$
\varepsilon (B^{\alpha}) = \varepsilon_{\alpha},\;\;\;
\varepsilon (C^{\alpha a}) = \varepsilon_{\alpha}+1.
$$
For reducible theories the complete set of field variables $\phi^A$
includes also pyramids of the ghosts, the antighosts and the Lagrange
multipliers which are combined into irreducible representations of the
$Sp(2)$--group (for more detailed discussion,  see [2]).

For
   each field $\phi^A$ of the total  configuration space
one introduces three kinds of antifields
$\phi^*_{Aa},\;\varepsilon(\phi^*_{Aa})=\varepsilon_A+1$
and $\bar\phi_A,\varepsilon(\bar\phi_A)=\varepsilon_A$. The antifields
$\phi^*_{Aa}$ maybe  treated as sources of BRST and antiBRST
transformations, while $\bar\phi_A$ corresponds to the source of their
combined transformation.

On the space of fields $\phi^A$ and antifields $\phi^*_{Aa}$ one
defines odd symplectic structures $(\;,\;)^a$ called the extended
antibrackets
$$
(F,G)^a\equiv\frac{\delta F}{\delta\phi^A}\;\frac{\delta
G}{\delta\phi^*_{Aa}}
-(F\leftrightarrow G)\;(-1)^{(\varepsilon(F)+1)(\varepsilon(G)+1)}
\eqno (3)
$$
The derivatives with respect to fields are understood as right and
those with respect to antifields as left.

The extended antibrackets have the following properties
$$
\varepsilon((F,G)^a)=\varepsilon(F)+\varepsilon(G)+1\;,
$$
$$
(F,G)^a=-(G,F)^a(-1)^{(\varepsilon(F)+1)(\varepsilon(G)+1)}\;,
$$
$$
(F,GH)^a=(F,G)^aH+(F,H)^aG(-1)^{\varepsilon(G)\varepsilon(H)}\;,
$$
$$
 {((F,G)^{\{a},H)^{b\}}(-1)^{(\varepsilon(F)+1)(\varepsilon(H)+1)}
+{\rm cycl.perm.} (F,G,H)}\equiv 0\;, \eqno(4)
$$
where curly brackets denote symmetrization with respect to the indices
$a,b$ of the $Sp(2)$ group. The last relations are the Jacobi identities
for the extended antibrackets. In particular, for any boson functional
$S,\;\varepsilon(S) =0$, one can establish that
$$
((S,S)^{\{a},S)^{b\}} \equiv 0
$$
The operators $V^a,\;\Delta^a$ are introduced as following
$$
V^a=\varepsilon ^{ab}\;\phi^*_{Ab}\;\frac
{\delta}{\delta\bar\phi_A},\;\varepsilon ^{ab}=-\varepsilon ^{ba},
\varepsilon ^{12}=1,\eqno (5)
$$
$$
\Delta^a=(-1)^{\varepsilon _A}\frac{\delta_{\it l}}{\delta\phi^A}
\frac{\delta}{\delta\phi^*_{Aa}},\eqno (6)
$$
Here the subscript $\it "l"$ denotes the left derivative with respect
to field. It maybe shown               that the algebra of operators
(5),(6) has the form
$$
V^{\{a}V^{b\}}=0,\; \Delta^{\{a}\Delta^{b\}}=0,\; \Delta^{\{a}V^{b\}}+
V^{\{a}\Delta^{b\}}=0.\;  \eqno (7)
$$
The action of the operators $V^a$ (4) upon the extended antibrackets is
given by the relations
$$
V^{\{a}(F,G)^{b\}} = (V^{\{a}F,G)^{b\}} - (-1)^{\varepsilon (F)}
(F,V^{\{a}G)^{b\}}. \eqno(8)
$$

The basic object of BLT scheme is a boson functional
$S=S(\phi,\phi^*_a,\bar\phi)$, satisfying the following generating
equations
$$
\frac{1}{2}(S,S)^a+V^aS=i\hbar\Delta^aS \eqno(9)
$$
with the boundary condition
$$
S|_{\phi^*_a = \bar\phi = \hbar = 0} = {\cal S}(A). \eqno (10)
$$
It should be noted that Eqs.(9) are compatible. The simplest way to
establish this fact is to rewrite Eqs.(9) in an equivalent form of
linear differential equations
$$
\bar\Delta^a \exp\left\{\frac{i}{\hbar}S\right\}=0.  \eqno(11)
$$
>From (7) it follows that the operators $\bar\Delta^a$ in (11) possess
the properties
$$
\bar\Delta^{\{a}\bar\Delta^{b\}}=0, \eqno (12)
$$
and therefore Eqs.(9) are compatible.

The quantum action $S_{\it ext}=S_{\it ext}(\phi,\phi^*_a,\bar\phi)$
for constructing of Feynman rules in the BLT scheme is introduced as
$$
\exp\left\{\frac{i}{\hbar}S_{\it ext}\right\}=\exp \left\{-i\hbar
\hat{T}(F)\right\}\exp\left\{\frac{i}{\hbar}S\right\}, \eqno(13)
$$
where the operator $\hat{T}$ has the form
$$
\hat{T}(F)=\frac{1}{2}\varepsilon_{ab}[\bar\Delta^b,[\bar\Delta^a,
F]_{-}]_{+}, \eqno(14)
$$
$F=F(\phi,\phi^*_a,\bar\phi)$ is the boson functional fixing a concrete
choice of admissible gauge. Note that the operator $\hat{T}$ commutes
with $\bar\Delta^a$ for arbitrary $F$
$$
[\hat{T},\bar\Delta^a]_{-}=0
\eqno(15)
$$
and hence $S_{\it ext}$ satisfies the equations (9)
$$
\bar\Delta^a \exp\left\{\frac{i}{\hbar}S_{\it ext}\right\}=0.
\eqno(16)
$$

The generating functional of the Green's functions $Z(J)$ is defined as
$$
Z(J) = {\cal Z}(J,\phi^*_a,\bar\phi)|_{\phi^*_a = \bar\phi = 0},
\eqno(17)
$$
where the extended generating functional ${\cal
Z}(J,\phi^*_a,\bar\phi)$ has the form
$$
{\cal Z}(J,\phi^*_a,\bar{\phi})=\int d\phi\exp\left\{\frac{i}{\hbar}
[S_{\it ext}(\phi,\phi^*_a,\bar{\phi})+J_A\phi^A]\right\}. \eqno(18)
$$
It is not difficult to show      that the integrand in Eq.(18) for
$J=0$ is invariant under the following global transformations
$$
\delta\phi^A = \frac{\delta S_{\it ext}}{\delta\phi^*_{Aa}}\mu_a,\;
\delta\phi^*_{Aa} = 0,\; \delta\bar{\phi}_A =
-\mu_a\varepsilon^{ab}\phi^*_{Ab}, \eqno (19)
$$
where $\mu_a$ is  the $Sp(2)$--doublet of constant Grassmann parameters
$(\varepsilon(\mu_a) = 1)$. These transformations are nothing but the
extended BRST ones in the BLT quantization.

Note that from the definitions (17),(18) and Eqs.(19) it follows that
the extended BRST transformations for the generating functional $Z(J)$
have the form
$$
\delta\phi^A = \left. \frac{\delta S_{\it
ext}}{\delta\phi^*_{Aa}}\right|_{\phi^* = \bar\phi = 0}\mu_a.
\eqno (20)
$$
The symmetry of the vacuum functional $Z(0)$ under the transformations
(20) permits to show the
                          independence of     $S$--matrix from the choice
of a gauge(within the BLT formalism [2].)

As a consequence of the fact that $S_{\it ext}$ satisfies  the
generating equations (16) one can write the
                                        Ward identities within the BLT
formalism. For the extended generating functional ${\cal Z}=
{\cal Z}(J,\phi^*_a, \bar\phi)$ (18) these identities have the form
$$
\bigg(J_A\frac{\delta}{\delta\phi^*_{Aa}} - \varepsilon^{ab}\phi^*_{Ab}
\frac{\delta}{\delta\bar{\phi}_A}\bigg){\cal Z} = 0. \eqno(21)
$$
Introducing in a standard manner the generating functional of the
vertex functions $\Gamma$,
$$
\Gamma(\phi,\phi^*_a,\bar{\phi}) = \frac{\hbar}{i}
\ln {\cal Z}(J,\phi^*_a,\bar\phi) - J_A\phi^A, \eqno(22)
$$
$$
\phi^A = \frac{\hbar}{i}\frac{\delta \ln {\cal Z}(J,\phi^*_a,\bar\phi)}
{\delta J_A},
$$
we obtain  the Ward identities
$$
\frac{1}{2}(\Gamma,\Gamma)^a + V^a\Gamma = 0 \eqno(23)
$$
These Ward identities will be used in the study of the gauge dependence
of EA.

\section{WARD IDENTITIES}
\hspace*{\parindent}

In this section we       derive the Ward identities for general
gauge theories with composite fields in the framework of BLT
covariant quantization.

Let us introduce the composite fields
$$
\sigma^m(\phi)=\sum_{n=2}\frac{1}{n!}\Lambda^m_{A_{1}...A_{n}}
\phi^{A_1}...\phi^{A_n},\;\;\;\;
\varepsilon(\sigma^m)\equiv \varepsilon_m.  \eqno (24)
$$
Using the quantum action $S_{\it ext}=
S_{\it ext}(\phi,\phi^*_a,\bar\phi)$ (13) we define the
generating functional $Z(J,\phi^*_a,\bar{\phi}, L)$ for the composite
fields $\sigma^m(\phi)$ as following
$$
Z(J,\phi^*_a,\bar{\phi},L)=\int
d\phi\exp\left\{\frac{i}{\hbar} [S_{\it
ext}(\phi,\phi^*_a,\bar{\phi})+J_A\phi^A+L_m\sigma^m(\phi)] \right \}=
$$
$$
=\exp\left\{\frac{i}{\hbar}W(J,\phi^*_a,\bar{\phi},L)\right \},
\eqno(24)
$$
where $W(J,\phi^*_a,\bar{\phi},L)$ is the generating functional of
the connected correlation functions for composite fields, and $L_m$
are sources for $\sigma^m$.

The Ward identities for general gauge theories with composite fields
are obtained as a consequence of the fact that $S_{\it ext}$ satisfies
the generating equations (16). To do this one can multiply Eqs.(16) on
functional
$$
\exp\left\{\frac{i}{\hbar}[J_A\phi^A+L_m\sigma^m(\phi)]\right \}
$$
and integrate over fields $\phi^A$. Then we have the identities
$$
\int d\phi\exp\left\{\frac{i}{\hbar}[J_A\phi^A+L_m\sigma^m(\phi)]
\right \}\bar\Delta^a\exp\left\{\frac{i}{\hbar}S_{\it ext}(\phi,
\phi^*_a,\bar{\phi})\right \}=0. \eqno(26)
$$
Carrying out integration by parts in (26) one derives the Ward
identities for functional $Z(J,\phi^*_a,\bar{\phi},L)$
$$
\left\{\Biggl( J_A+L_m\sigma^m_{,A}\Biggl(\frac{\hbar}{i}
\frac{\delta}{\delta J}\Biggr)\Biggr)\frac{\delta}
{\delta\phi^*_{Aa}}-V^a\right\}Z(J,\phi^*_a,\bar{\phi},L)=0.
\eqno(27)
$$
For the functional $W(\phi,\phi^*_a,\bar{\phi},L)$ the identities
(27) are
$$
\left\{\Biggl( J_A+L_m\sigma^a_{,A}\Biggl(\frac{\delta W}{\delta J}
+\frac{\hbar}{i}\frac{\delta}{\delta J}\Biggr)\Biggr)\frac{\delta}
{\delta\phi^*_{Aa}}-V^a\right\}W(J,\phi^*_a,\bar{\phi},L)=0.
\eqno(28)
$$
Here we have used the notations
$$
\sigma^m_{,A}(\phi)\equiv\frac{\delta \sigma^m(\phi)}{\delta\phi^A}.
$$

Let us introduce the generating functional of vertex functions
(effective action) for composite fields $\Gamma=\Gamma(\phi,\phi^*_a,
\bar{\phi},\Sigma)$ by the rule
$$
\Gamma(\phi,\phi^*_a,\bar{\phi},\Sigma)=W(J,\phi^*_a,\bar{\phi},L)-
J_A\phi^A-L_m(\Sigma^m+\sigma^m(\phi)), \eqno(29)
$$
where
$$
\phi^A=\frac{\delta W(J,\phi^*_a,\bar{\phi},L)}{\delta J_A},
$$
$$
\Sigma^m=\frac{\delta W(J,\phi^*_a,\bar{\phi},L)}{\delta L_m}-
\sigma^m\Biggl(\frac{\delta W}{\delta J}\Biggr). \eqno(30)
$$
>From definitions (29),(30) it follows that
$$
\frac{\delta\Gamma}{\delta\phi^A}=- J_A -
L_m\sigma^m_{,A}\Biggl(\frac{\delta W}{\delta J}\Biggr),
\frac{\delta\Gamma}{\delta\Sigma^m}=-L_m \eqno (31)
$$
Using expressions (28) and definition (29) one can obtain the
Ward identities for $\Gamma(\phi,\phi^{*}_a,\bar{\phi},\Sigma)$
in the form
$$
\frac{1}{2}(\Gamma,\Gamma)^a+V^a\Gamma+\frac{\delta\Gamma}
{\delta\Sigma^m}\Biggl(\sigma^m_{,A}({\hat \phi})-\sigma^m_{,A}(\phi)
\Biggr)\frac{\delta\Gamma}{\delta\phi^{*}_{Aa}}=0 , \eqno(32)
$$
where
$$
\hat{\phi}^A=\phi^A+i\hbar (G^{''-1})^{A\alpha}\frac{\delta_{\it l}}
{\delta\Phi^{\alpha}},\eqno(33)
$$
$$
\Phi^{\alpha}=(\phi^A,\Sigma^m),\;\;G^{''}_{\alpha\beta}=\frac{\delta_
{\it l}E_{\beta}}{\delta\Phi^{\alpha}}, \eqno(34)
$$
$$
E_{\alpha}=\Biggl(\frac{\delta\Gamma}{\delta\phi^A}-
\frac{\delta\Gamma}{\delta\Sigma^m}\sigma^m_{,A}(\phi),
\frac{\delta\Gamma}{\delta\Sigma^m}\Biggr). \eqno(35)
$$
These identities are useful in various aspects, in particularly,
in the study of the gauge dependence. They generalize the
corresponding Ward identities of Section 2 for the case of
composite fields.

\section{THE GAUGE DEPENDENCE}
\hspace*{\parindent}

In this section we discuss        the gauge dependence of
generating functionals $Z,W,\Gamma$ for general gauge theories
with composite fields. The derivation of this dependence is based
on the fact that any variation of gauge functional $F\rightarrow
F+\delta F$ leads to variation of action $S_{\it ext}$ (13) and
functional $Z$ (25). One can easily check that the variation of
action can be expressed in the form
$$
\delta \Biggl(\exp\left\{\frac {i}{\hbar}S_{\it ext}\right\}\Biggr)=-
i\hbar\hat{T}(\delta\hat{X})\exp\Biggl\{\frac{i}{\hbar}
S_{\it ext}\Biggr\} \eqno(36)
$$
with some operator $\delta\hat{X}$ of first order with respect to
$\delta F$. For our purposes it is not important to know explicit
expression of operator $\delta\hat{X}$ through $\delta F$ . Note
only that one can always present the operator $\delta\hat{X}$ in
the following way
$$
\delta\hat{X}\Biggl(\phi,\phi^*_a,\bar\phi{;}\frac{\delta}{\delta\phi},
\frac{\delta}{\delta\phi^*_a}, \frac{\delta}{\delta\bar\phi},\Biggr)
=\sum_{n,m,l=0}\Biggl(\delta X^{A_1...A_n}_{B_1b_1...B_mb_m
C_1...C_l}(\phi,\phi^*_a,\bar{\phi})
$$
$$
\frac{\delta}{\delta\phi^{A_1}}...\frac{\delta}{\delta\phi^{A_n}}
\frac{\delta}{\delta\phi^*_{B_1b_1}}...\frac{\delta}{\delta\phi^*_
{B_mb_m}}
\frac{\delta}{\delta\bar{\phi}_{C_1}}...\frac{\delta}{\delta
\bar{\phi}_{C_l}}\Biggr). \eqno(37)
$$
>From Eqs.(13), (15), (26) it follows that variation of functional
$Z$ can be written as
$$
\delta Z(J,\phi^*_a,\bar{\phi},L)=
$$
$$
=\int d\phi\exp\left\{\frac{i}{\hbar}[J_A\phi^A+
L_m \sigma^{m}(\phi)]\right\}
(-i\hbar\hat{T}(\delta\hat{X}))\exp\left\{\frac{i}{\hbar}
S_{\it ext}(\phi, \phi^*_a,\bar{\phi})\right \}=
$$
$$
=-\frac {i\hbar}{2} \varepsilon_{ab}
\int d\phi\exp\left\{\frac {i}{\hbar}[J_A\phi^A+
L_m\sigma^m(\phi)] \right\}
\bar{\Delta}^b\bar{\Delta}^a\delta\hat{X}\exp\left\{
\frac{i}{\hbar}S_{\it ext}(\phi, \phi^*_a,\bar{\phi})\right \}.
\eqno(38)
$$
Carrying out integration by parts in the functional integral (38)
one can rewrite the variation of $Z$ in the form
$$
\delta Z(J,\phi^*_a,\bar{\phi},L)=
$$
$$
=\frac {i}{2\hbar} \varepsilon_{ab}{\hat q}^b{\hat q}^a
\delta\hat{X}\Biggl(\frac{\hbar}{i}\frac{\delta}{\delta J},\phi^*_a,
\bar\phi{;} \frac {1}{i\hbar}\biggl(J+L\sigma_,\biggl(
\frac {\hbar\delta}{i\delta J}\biggr)\biggr),\frac{\delta}{\delta\phi^*_a}
,\frac{\delta}{\delta\bar\phi}\Biggr)Z(J,\phi^*_a,\bar{\phi},L),
\eqno(39)
$$
where ${\hat q}^a$ stands for an operatorial $Sp$(2)-doublet
$$
{\hat
q}^a=-\biggl[J_A+L_m\sigma^m_{,A}\biggl(\frac{\hbar}{i}\frac{\delta}
{\delta J}\biggr)\biggr] \frac{\delta}{\delta\phi^*_{Aa}}+V^a\;,
\eqno(40)
$$
which is directly verified to satisfy the relations
$$
{\hat q}^{\{a}{\hat q}^{b\}}=0
\eqno(41)
$$

             From the relations (39),(40) taking into account of the
Ward identities (27) for $Z$ and the fact that
$$
\delta
Z=\frac{i}{\hbar}\delta W\;Z\;,
$$
it follows the expression for the
variation of functional $W(J,\phi^{*}_{a},{\bar\phi},L)$:
$$
\delta W(J,\phi^{*}_a ,{\bar
\phi},L)=\frac{1}{2}\varepsilon_{ab}{\hat Q}^b{\hat Q}^a<\delta {\hat
X}>\;,
\eqno(42)
$$
where the operators ${\hat Q}^a$ are related to
${\hat q}^a$ through a unitary transformation
$$
{\hat Q}^a={\rm exp}\{-\frac {i}{\hbar}W\}{\hat q}^a
{\rm exp}\{\frac {i}{\hbar}W\}
$$
and have the form
$$
{\hat Q}^{a}=-\biggl[J_A+L_m\sigma^m_{,A}\biggl(\frac{\delta
W}{\delta J}+\frac{\hbar}{i}\frac{\delta} {\delta J}\biggr)\biggr]
\frac{\delta}{\delta\phi^*_{Aa}}+V^{a} \;.
\eqno(43)
$$
As a consequence of the Eq.(41) ${\hat R}^{a}$ possess the following
properties
$$
{\hat Q}^{\{a}{\hat Q}^{b\}}=0\;.
\eqno(44)
$$
 In (42), the notation $<\delta {\hat X}>$ is used for the
vacuum expectation value of the operator $\delta {\hat X}$
$$
<\delta {\hat X}>=
$$
$$
\delta {\hat X}\Biggl(\frac {\delta W}{\delta J} + \frac
{\hbar\delta}{i\delta J},\phi^{*}_{a} ,{\bar \phi}{;}  \frac
{1}{i\hbar}\biggl(J+L\sigma_,\biggl(\frac {\delta W}{\delta J} +\frac
{\hbar\delta}{i\delta J}\biggr)\biggr),\frac{\delta}{\delta\phi^*_a}
+\frac{i}{\hbar}\frac{\delta W}{\delta\phi^*_a},\frac{\delta}
{\delta\bar\phi}+\frac{i}{\hbar}\frac{\delta
W}{\delta\bar\phi}\Biggr)\;.
$$

Let us find the expression for $\delta\Gamma(\phi,\phi^{*}_{a},
{\bar\phi},\Sigma)$. To this end, we must study some differential
consequences from the Ward identities for $Z$, $W$ and use the
following observations that are a consequence of definitions
(29)-(31).  Namely,
$$
\delta W=\delta \Gamma\;,
$$
$$
{\frac{\delta}{\delta\phi^*_a}}_{\mid
J,L}={\frac{\delta}{\delta\phi^*_a}}_{\mid
\phi,\Sigma} + \frac{\delta
\phi}{\delta\phi^*_a} {\frac{\delta_{\it l}} {\delta\phi}}_{\mid
\phi^{*}_{a},\Sigma} + \frac{\delta
\Sigma}{\delta\phi^*_a} {\frac{\delta_{\it l}}{\delta \Sigma}}_{\mid
\phi,\phi^{*}_{a}}\;,
$$
$$
V^{a}_{\mid J,L}=V^{a}_{\mid
\phi,\Sigma} + {V^{a}\phi \frac{\delta_{\it
l}}{\delta\phi}}_{\mid {\bar \phi},\Sigma} + {V^{a}\Sigma
\frac{\delta_{\it l}}{\delta \Sigma}}_{\mid \phi,{\bar
\phi}}\;.
\eqno(45)
$$
Next, differentiating the Ward identities for $Z$ (27) with respect
to the sources $J$ and $L$, then rewriting these relations for the
functional $W$ and transforming the latter with allowance for
Eqs.(29)-(31) we obtain
$$
{\hat Q}^{a}\phi^{A}_{\mid L,J}=
\frac {\delta \Gamma}{\delta\phi^{*}_{Aa}}(-1)^{\varepsilon_A}+
$$
$$
\frac {i}{\hbar}\Biggl(\phi^{A}\frac
{\delta\Gamma}{\delta\Sigma^{m}}\sigma^{m}_{,B}({\hat \phi}) \frac
{\delta \Gamma}{\delta \phi^{*}_{Ba}}(-1)^{\varepsilon_A}- \frac
{\delta\Gamma}{\delta\Sigma^{m}}\sigma^{m}_{,B}({\hat \phi}) \frac
{\delta \Gamma}{\delta \phi^{*}_{Ba}}\phi^{A}\Biggr)\;,
$$
$$
{\hat Q}^a\Sigma^{n}_{\mid L,J}=
\biggl(\sigma^{n}_{,B}({\hat \phi})-
\sigma^{n}_{,B}(\phi)\biggr)\frac{\delta\Gamma}{\delta\phi^{*}_{Ba}}
(-1)^{\varepsilon_n}+ \;.
$$
$$
\frac {i}{\hbar}\Biggl(\biggl(\Sigma^{n}+\sigma^{n}(\phi)\biggr)\frac
{\delta\Gamma}{\delta\Sigma^{m}}\sigma^{m}_{,B}({\hat \phi}) \frac
{\delta \Gamma}{\delta \phi^{*}_{Ba}}(-1)^{\varepsilon_n}-
$$
$$
\frac{\delta\Gamma}{\delta\Sigma^{m}}\sigma^{m}_{,B}({\hat \phi}) \frac
{\delta \Gamma}{\delta
\phi^{*}_{Ba}}\biggl(\Sigma^{n}+\sigma^{n}(\phi)\biggr)\Biggr)-
$$
$$
\frac
{i}{\hbar}(-1)^{\varepsilon_n}\sigma^{n}_{,C}(\phi)\Biggl(\phi^{C}\frac
{\delta\Gamma}{\delta\Sigma^{m}}\sigma^{m}_{,B}({\hat \phi}) \frac
{\delta \Gamma}{\delta \phi^{*}_{Ba}}-(-1)^{\varepsilon_C}\frac
{\delta\Gamma}{\delta\Sigma^{m}}\sigma^{m}_{,B}({\hat \phi}) \frac
{\delta \Gamma}{\delta \phi^{*}_{Ba}}\phi^{C}\Biggr)+
$$
$$
(-1)^{\varepsilon_n}\Biggl(\sigma^{n}_{,C}(\phi)\frac
{\delta\Gamma}{\delta\Sigma^{m}}\sigma^{m}_{,B}({\hat \phi})
(G^{''-1})^{C\alpha}\frac{\delta_{\it l}}{\delta \Phi^\alpha} \frac
{\delta \Gamma}{\delta
\phi^{*}_{Ba}}(-1)^{\varepsilon_{C}\varepsilon_{B}}-
$$
$$
(-1)^{\varepsilon_{n}\varepsilon_{B}}\frac
{\delta\Gamma}{\delta\Sigma^{m}}\sigma^{m}_{,B}({\hat \phi})
\sigma^{n}_{,C}(\phi)(G^{''-1})^{C\alpha}\frac{\delta_{l}}
{\delta\Phi^{\alpha}} \frac {\delta \Gamma}{\delta
\phi^{*}_{Ba}}\Biggr)\;.
\eqno(46)
$$
Taking into account the relations (43),(45)-(46), we have the final
representation for the variation of the effective action with composite
fields
$$
\delta\Gamma(\phi,\phi^{*}_{a},{\bar \phi},\Sigma)= \frac
{1}{2}\varepsilon_{ab}{\hat s}^{b}{\hat s}^{a}<<\delta {\hat X}>> \;,
\eqno(47)
$$
where we have introduced the notations
$$
{\hat s}^{a}=(\Gamma,\;\;)^{a}+V^{a}+(-1)^{\varepsilon_m}\Biggl(\biggl(
\sigma^{m}_{,A}({\hat \phi})- \sigma^{m}_{,A}(\phi)\biggr)\frac {\delta
\Gamma}{\delta \phi^{*}_{Aa}}\Biggr)\frac {\delta_{\it l}}{\delta
\Sigma^{m}} + \frac
{\delta\Gamma}{\delta\Sigma^{m}}\biggl(\sigma^{m}_{,A}({\hat\phi})
-\sigma^{m}_{,A}(\phi)\biggr)\frac {\delta}{\delta
\phi^{*}_{Aa}}-
$$
$$
\frac {i}{\hbar}\Biggl[\frac
{\delta\Gamma}{\delta\Sigma^{m}}\Biggl(\sigma^{m}_{,B}({\hat \phi})
\frac {\delta \Gamma}{\delta \phi^{*}_{Ba}}\Phi^{\alpha}\Biggr)-
(-1)^{\varepsilon_{\alpha}}\Phi^{\alpha}
\frac {\delta\Gamma}{\delta\Sigma^{m}}\Biggl(\sigma^{m}_{,B}({\hat \phi})
\frac {\delta \Gamma}{\delta \phi^{*}_{Ba}}\Biggr)\Biggr]\frac
{\delta_{\it l}}{\delta \Phi^{\alpha}}-
$$
$$
\frac {i}{\hbar}\Biggl[\frac
{\delta\Gamma}{\delta\Sigma^{m}}\Biggl(\sigma^{m}_{,B}({\hat \phi})
\frac {\delta \Gamma}{\delta
\phi^{*}_{Ba}}\sigma^{n}(\phi)\Biggr)-(-1)^{\varepsilon_n}\sigma^{n}(\phi)
\frac {\delta\Gamma}{\delta\Sigma^{m}}\Biggl(\sigma^{m}_{,B}({\hat \phi})
\frac {\delta \Gamma}{\delta \phi^{*}_{Ba}}\Biggr)\Biggr]\frac
{\delta_{\it l}}{\delta \Sigma^{n}}+
$$
$$
\frac {i}{\hbar}(-1)^{\varepsilon_n}\sigma^{n}_{,C}(\phi)\Biggl[\frac
{\delta\Gamma}{\delta\Sigma^{m}}\Biggl(\sigma^{m}_{,B}({\hat \phi})
\frac {\delta \Gamma}{\delta
\phi^{*}_{Ba}}\phi^{C}\Biggr)(-1)^{\varepsilon_{C}}
- \phi^{C} \frac
{\delta\Gamma}{\delta\Sigma^{m}}\Biggl(\sigma^{m}_{,B}({\hat \phi})
\frac {\delta \Gamma}{\delta \phi^{*}_{Ba}}\Biggr)\Biggr]\frac
{\delta_{\it l}}{\delta \Sigma^{n}}+
$$
$$
(-1)^{\varepsilon_m}\Biggl[\sigma^{m}_{,C}(\phi)\frac {\delta\Gamma}
{\delta\Sigma^n}\Biggl(\sigma^{n}_{,A}({\hat\phi})(G^{''-1}
)^{C\alpha}\frac {\delta_{\it l}}{\delta\Phi^{\alpha}}\frac{\delta
\Gamma}{\delta \phi^{*}_{Aa}}\Biggr)(-1)^{\varepsilon_{C}\varepsilon_A}-
$$
$$
(-1)^{\varepsilon_{m}\varepsilon_A}\frac{\delta\Gamma}
{\delta\Sigma^n}\Biggl(\sigma^{n}_{,A}({\hat\phi})\sigma^{m}_{,C}
(\phi)(G^{''-1})^{C\alpha} \frac
{\delta_{\it l}}{\delta\Phi^{\alpha}}\frac{\delta \Gamma}{\delta
\phi^{*}_{Aa}}\Biggr)\Biggr]\frac {\delta_{\it l}}{\delta \Sigma^m} \;,
\eqno(48)
$$
$$
<<\delta {\hat X}>>\equiv
$$
$$
\equiv\delta {\hat X}\Biggl({\hat
\phi},\phi^{*}_{a},{\bar \phi};\frac
{1}{i\hbar}\biggl(-\Gamma_{,} + \Gamma_{,m}\biggl[\sigma^{m}_{,}(\phi) -
\sigma^{m}_{,}({\hat\phi})\biggr]\biggr),
\;\;\;\frac{\delta}{\delta\phi^*_a}+\frac{i\delta
\Gamma}{\hbar\delta\phi^*_a}
,\frac{\delta}{\delta\bar\phi}+\frac{i\delta
\Gamma}{\hbar\delta\bar\phi}\Biggr)\;,
\eqno(49)
$$
while the action of the operators $\biggl[(\Gamma,\;\;)^{a}+V^{a}\biggr]$
on an arbitrary functional $G=G(\phi,\phi^{*}_{a},{\bar\phi},\Sigma)$,
 is understood as follows:
$$
\biggl[(\Gamma ,\;\;)^{a} + V^{a}\biggr]G
\equiv (\Gamma,G)^{a} + V^{a}G\equiv {\hat s}^{a}_{0}G\;.  \eqno(50)
$$

The operators ${\hat s}^{a}_{0}$ are said to be the generators of the
quantum  extended BRST transformations without composite fields in
the BLT method for general gauge theories, while the operators ${\hat
s}^{a}$ can be considered as a deformation of generators
${\hat s}^{a}_{0}$ related to the presence of the composite fields
in the theory.

Since the values ${\hat s}^{a}$ are obtained through a change of
variables, i.e. the Legendre transformation, from the operatorial
Sp(2)-doublet ${\hat Q}^a$ satisfying Eq.(44), the operators ${\hat
s}^{a}$ must possess the following algebra
$$
{\hat s}^{a}{\hat s}^{b}=0
\eqno(51)
$$
This is the standard algebra for the extended BRST symmetry. It is
surprising that this algebra remains the same in case of an arbitrary
composite field. Note,however, that the above operators are defined
with the help of composite fields EA which satisfy the Ward
identities.

The corresponding algebra of the operators ${\hat s}^{a}_{0}$ has the
standard form as

\noindent
for ${\hat s}^{a}$
$$
{\hat s}^{\{a}_{0}{\hat s}^{b\}}_{0}=0\;,
\eqno(52)
$$
which is a consequences of the algebra of operators $V^a$ in (7), of
Jacobi identities for the extended antibrackets (4), of the Leibnitz
rule of the action of the $V^a$ on the extended antibrackets (8) and of
the Ward identities for $\Gamma(\phi,\phi^{*}_{a},{\bar\phi})$ (23).

It should be noted that the expression for $\delta\Gamma$ (47) has very
simple and remarkable form.  It is defined through the commutator of
the operators ${\hat s}^{a}$. We expect that
the gauge dependence of general gauge theories can be understood also
in geometrical terms using, for example, the local BRST cogomology in
the same style as in refs.[28].

Note                              that the expression for $\delta\Gamma$
can be rewritten in the following equivalent form (cf. Ref.5) with the help
of systematic use of Ward identities for $\Gamma$ (32) and their
differential consequences:
$$
\delta\Gamma(\phi,\phi^{*}_{a},{\bar
\phi},\Sigma)=\frac {\delta \Gamma}{\delta \Phi^{\alpha}}W^{\alpha} +
\phi^{*}_{Aa}D^{Aa}
\eqno(53)
$$
with the fully definite functionals
$W^{\alpha}$ and $D^{Aa}$ depending   on all the variables
$\phi,\phi^{*}_{a},{\bar \phi},\Sigma$.

>From Eq.(53) we get the following Theorem:
                              the generating functional of vertex
functions $\Gamma(\phi,\phi^{*}_{a},{\bar \phi},\Sigma)$ in the BLT
quantization method of general gauge theories with composite fields
does not depend   on the gauge on its extremals which are defined as
$$
\frac{\delta\Gamma}{\delta\Phi^{\alpha}} = 0 \eqno(54)
$$
and the hypersurface defined by conditions
$$
\phi^{*}_{Aa}=0.  \eqno(55)
$$
It is useful to compare this result with the gauge dependence of
generating functional of vertex functions with composite fields in the
BV--quantization method. This problem has been investigated in ref.[22]
with the following result-
                                            generating functional
of vertex functions with composite fields is gauge independent only
on its extremals.Here
we have a more complicated case.

With the help of Eq.(52) one can develop
                          another point of view on the problem of
gauge dependence. Namely, the variation of effective action in the BLT
quantization method for
                       general gauge theories with composite fields is
proportional to the commutator of the operators ${\hat s}^a$
 which act on the corresponding variation $<<\delta{\hat X}>>$.

Notice that an essential feature of the our proof   is the
assumption of existence of "deep" gauge invariant regularization
preserving the Ward identities and permitting to use their differential
consequences. Then we expect that
                       the corresponding completely renormalized generating
functionals satisfy                             the same properties as the
nonrenormalizable ones.

\section{APPLICATIONS: INDUCED GRAVITATIONAL - VECTOR INTERACTION}
\hspace*{\parindent}

Let us discuss some cosmological application of
                                   the composite fields effective
action in the external gravitational field.             Note that the
BLT formalism still has not been generalized to the problems with
external fields, so this section lies outside of the general
study developed in this paper.

There were some discussions recently on the presence of magnetic field
in the early Universe. It could be of a primordial origin, and be
produced in the inflationary Universe. However with usual Maxwell -
type lagrangian it seems to be impossible to produce such the magnetic
field [23]. One can consider magnetic field in string cosmology [24], or
by addition the terms of form $RA_{\mu}A^{\mu}$ to Maxwell lagrangian.
However, such terms break gauge invariance.

The way out maybe found by using the composite fields effective action.
Let us consider the generating functional $W$ for Maxwell theory ($\hbar
= 1$):
$$
\exp\{i\;W[J,K]\} = \int {\cal D}A^a_{\mu}\exp{\{i[S + \int d^4x\sqrt{-g}(J_{\mu}
A^{\mu} + KA_{\mu}A^{\mu})]\}} \eqno(56)
$$
where $S = -\frac{1}{4}\int d^4x\sqrt{-g}F_{\mu\nu}F^{\mu\nu}$, the
Landau gauge which is
an effective gauge is chosen, the presence of the corresponding gauge
breaking term and ghost action is supposed.
We don't discuss the renormalization of the
vacuum sector (action for external gravitational field ) which has been
studied   in all detail in [25].

Study of the generating functional (56) shows that its renormalization
induces few terms. Among of them     there is term of the form $KR$
[26]. Using the correspondent non-homogeneous renormalization group
equation and explicit one-loop calculations one can show that Maxwell
sector is modifying as
$$
W[0,K] = -\frac{1}{4}\int d^4x\sqrt{-g} \biggl\{F_{\mu\nu}F^{\mu\nu} -
\frac{1}{(4\pi)^2}KR
\ln \frac{|R+K|}{{\mu}^2} - \frac{12K^2}{(4\pi)^2} + ...\biggr\} .
$$
Hence, the gravitational-vector interaction is induced on quantum level.
Such a term maybe relevant to produce the strong enough magnetic field
in the early  Universe in a consistent way (without breaking of gauge
invariance on classical level). Hence, the development of quantization
schemes for theories in the external fields is getting quite important.

\section{DISCUSSION}
\hspace*{\parindent}

In summary, composite fields EA for general gauge theories
is investigated in frames of BLT-quantization method (Sp(2)
formalism). The new set of operators which depend on composite fields
is introduced. Their algebra coincides with the algebra of extended
BRST transformations. The variation of composite fields EA is found
in a very simple form, using this new set of operators and Ward
identities. The proof of on-shell gauge fixing independence of the
composite fields EA is given. Some properties of EA for composite
vector fields are briefly discussed.

The importance of the composite fields EA in external background
has been mentioned recently (see for example [27]) in connection
with the study of the exact potential in supersymmetric YM theories.
Hence, it is necessary to generalize the quantization methods
existing already also for composite fields EA in an external background.
>From another side, the connected study of the structure and properties
of gauge theories in the presence of external fields
 is getting of interest due to possible applications.
Such an investigation maybe quite non-trivial what follows
from results of ref.[30] where Ward identities for gauge theories
have been obtained
within BV quantization [1] in the presence of external fields.
Hence, it is very interesting to generalize the results of this work
(Ward identities, Theorem) for the case of external gauge
and (or) gravitational background.

Another interesting line of research is related with the study
of BRST cohomologies a la ref.[28] in BLT-formalism. Recently,
using results of [28] it has been shown that non-renormalizable
theories maybe renormalizable in a modern sense [29] (taking
into account the infinite number of counterterms). Then it would
be of interest to formulate the proof of ref.[29] within
BLT formalism (even taking into account composite fields).

\vspace{2mm}
\noindent{\bf Acknowledgments}
We would like to thank J.Gomis and I.V.Tyutin for useful discussions
 of the paper.

\vspace{.5mm}

\section{APPENDIX A}
\hspace*{\parindent}

In this Appendix we will give few remarks on Wilson action for
composite fermion fields. We consider the theory containing spinors
$\psi$ and gauge fields $A^a_{\mu}$.

The generating functional $W$ (euclidean notations are used) is defined
as
$$
\exp (-W[J]) = \int {\cal D}\phi\exp\{-S_L(\psi,\bar{\psi},A_{\mu})
-J\bar{\psi}\psi\}, \eqno (A1)
$$
where $\phi$ is the set of all fields (including ghost) and ghost term
and gauge -fixing term are supposed to be present in (A1). The Wilson
effective action is defined through the introduction of the infra-red
cut-off $L$, and background field propagators in (A1) are modified
(compare with standard) as
$$
K^{-1}_L = \int_{0}^{L} dt \exp(-tK) \eqno (A2)
$$
in the action $S_L$ (for $L\rightarrow {\infty}$ it becomes the
standard propagator). The corresponding flow equation for $W$ maybe
written.

The Wilson effective action is defined via the Legendre transform
$$
\Gamma_L(<\bar{\psi}\psi >) = W[J] - J<\bar{\psi}\psi > \eqno(A3)
$$
Such Wilson effective action for composite fields maybe easily applied
to study the   exact results in SUSY theories (for more details, see [27]).
It would be extremely interesting to combine BLT-formalism with Wilson
effective action formalism. That would definitely enrich both
approaches, but requires quite hard work.

\newpage
\begin{center}
{\bf References}
\end{center}
\medskip
\begin{itemize}

\item[{\hfill [1]}]
I.A.Batalin and G.A.Vilkovisky,
Phys.Lett.,{\bf B102}, 27 (1981);  {\it ibid} {\bf B120}, 166
(1983); Phys.Rev., {\bf D28}, 2567 (1983); Nucl.Phys., {\bf B234},106
(1984); J.Math.Phys., {\bf 26}, 172 (1985).

\item[{\hfill [2]}]
I.A.Batalin , P.M.Lavrov and I.V.Tyutin ,
J.Math. Phys.,{\bf 31}, 1487 (1990); {\it ibid} {\bf 32}, 532 (1991);
{\it ibid} {\bf 32}, 2513 (1991).

\item[{\hfill [3]}]
M.Henneaux and C.Teitelboim, Quantization of Gauge Systems,
Princeton NJ, Princeton Press, 1992.

\item[{\hfill [4]}]
J.Gomis,J.Paris and S.Samuel, Phys.Repts.,{\bf 259}, 1 (1995)

\item[{\hfill [5]}]
P.M.Lavrov, Mod.Phys.Lett.,{\bf A6}, 2051 (1991)

\item[{\hfill [6]}]
M.Henneaux, Phys.Lett.,{\bf B282}, 372 (1992).

\item[{\hfill [7]}]
G.Barnich, R.Conctantinescu and P.Gregoire, Phys.Lett.,
{\bf B293}, 353\\ (1992).

\item[{\hfill [8]}]
P.Gregoire and M.Henneaux, J.Phys.A:Math.Gen., {\bf 26}, 6073
(1993).

\item[{\hfill [9]}]
P.H.Damgaard and F.De Jongle, Phys.Lett., {\bf B305}, 59 (1993).

\item[{\hfill [10]}]
A.Nersessian and P.H.Damgaard , Phys.Lett., {\bf B355}, 150 (1995).

\item[{\hfill [11]}]
P.H.Damgaard ,F.De Jongle and K.Bering, Nucl.Phys., {\bf B455}, 440
(1995).

\item[{\hfill [12]}]
I.Batalin and R.Marnelius, Phys.Lett., {\bf B305}, 44 (1995).

\item[{\hfill [13]}]
L.T\u{a}taru, J.Phys.A:Math.Gen., {\bf 28}, 4175 (1995).

\item[{\hfill [14]}]
P.M.Lavrov and A.A.Reshetnyak, Phys.Atom.Nucl., {\bf 58} 324 (1995).

\item[{\hfill [15]}]
P.M.Lavrov and P.Yu.Moshin, Theor.Math.Phys., {\bf 102}, 60 (1995).

\item[{\hfill [16]}]
P.M.Lavrov, Phys.Lett., {\bf B366}, 160 (1996)

\item[{\hfill [17]}]
J.M.Cornwall, R.Jackiw and E.Tomboulis, Phys.Rev., {\bf D10}, 2428
(1974).

\item[{\hfill [18]}]
R.W.Haymaker, Rivista Nuovo Cim., {\bf 14}, 1 (1991).

\item[{\hfill [19]}]
Y.Nambu and G.Jona-Lasinio, Phys.Rev., {\bf 122}, 345 (1961);
D.Gross and A.Neveu,
Phys.Rev., {\bf D10}, 3235 (1974).

\item[{\hfill [20]}]
W.Bardeen, C.Hill and M.Lindner, Phys.Rev., {\bf D41}, 1647 (1990).

\item[{\hfill [21]}]
N.Seiberg, hep-th 9506077

\item[{\hfill [22]}]
P.M.Lavrov and S.D.Odintsov, Int.J.Mod.Phys.A, {\bf 4}, 5205
 (1989)

\item[{\hfill [23]}]
M.S.Turner and L.M.Widrow, Phys.Rev., {\bf D37}, 2743 (1987).

\item[{\hfill [24]}]
M.Gasperini,M.Giovannini and G.Veneziano, Phys.Rev.Lett, {\bf 75},
3796 (1995).

\item[{\hfill [25]}]
I.L.Buchbinder,S.D.Odintsov and I.L.Shapiro, Effective Action in
Quantum Gravity, IOP Publishing, Bristol and Philadelphia, 1992.

\item[{\hfill [26]}]
I.L.Buchbinder and S.D.Odintsov, Europhys.Lett., {\bf 4} 149 (1987).

\item[{\hfill [27]}]
S.P.de Alwis, preprint COLD-HEP-363, 1995.

\item[{\hfill [28]}]
G.Barnich,F.Brandt and M.Henneaux, Comm.Math.Phys.,{\bf 174} 57
(1995);{\bf 174} 93 (1995).

\item[{\hfill [29]}]
J.Gomis and S.Weinberg, preprint UTTG-18-95.

\item[{\hfill [30]}]
F.Bastianelli, Phys.Lett. {\bf B263} 411 (1991).

\end{itemize}

\end{document}